\def\pr{Phys. Rev. }
\documentclass[ twocolumn,amsmath, amssymb,prl]{revtex4}
\usepackage{hyperref}

\usepackage{graphicx}
\usepackage{amssymb}

\begin{document}
\title{Electron beam formation from spin-orbit interactions in zincblende semiconductor quantum wells} %
\author{David H. Berman}
\author{Michael E. Flatt{\'e}}
\affiliation{Department of Physics and Astronomy, University of Iowa, Iowa City, IA 52242}

\begin{abstract}
We find a dramatic enhancement of electron propagation along a narrow range of real-space angles from an isotropic source in a two-dimensional quantum well made from a zincblende semiconductor. This ``electron beam'' formation is caused by the interplay between spin-orbit interaction originating from a perpendicular electric field to the quantum well and the intrinsic spin-orbit field of the zincblende crystal lattice in a quantum well, in situations where the two fields are {\it different in strength} but of the same order of magnitude.  Beam formation is associated with caustics and can be described semi-classically using a stationary phase analysis.  
\end{abstract}
\maketitle


Spin transport in semiconductors\cite{Wolf2001,Awschalom2002,Ziese2001,Awschalom2007} can be dramatically modified by spin-orbit interactions, producing such effects as  coherent precession without a magnetic field\cite{Schliemann2003,Kato2004c,Crooker2005a,Crooker2005b,Bernevig2006,Koralek2009}. These effects vanish, however, for transport of unpolarized electrons, as the features generated for an initial electron spin polarized up are complementary to those generated from an initial electron with spin down. Spin-orbit effects have also been found to generate spontaneous spin polarization\cite{Edelstein1990} and spin currents\cite{Dyakonov1971,Hirsch1999,Murakami2003,Sinova2004,Kato2004b} in the presence of the flow of unpolarized electrons.    An open question is whether the {\it spin-averaged transport of  unpolarized} spins can be influenced by the spin-orbit interaction in a non-trivial fashion. The outward flux of electrons injected at a point in a quantum well is isotropic (albeit with interesting spin structure), in the presence of the spin-orbit field from a  perpendicular electric field\cite{Walls2006,Csordas2006} (Rashba effect\cite{Rashba1960,Bychkov1984}) or from the zincblende lattice of the constituent semiconductors\cite{Bruning2007} (Dresselhaus field\cite{Dresselhaus1955,Awschalom2002}), or when the two spin-orbit fields are of the same strength\cite{Bernevig2006,Koralek2009}. Calculations of transport in the mixed Rashba and Dresselhaus system based the Boltzmann equation\cite{Trushin2007,Trushin2009} find isotropic conductivity for nonmagnetic systems, however these neglect interference between states of different momenta.

Here we describe a dramatic enhancement of electronic propagation along a narrow range of real-space angles which occurs in the presence of  Rashba and Dresselhaus spin-orbit fields of specific, different strengths. The angular width of this ``electron beam'' depends sensitively on the ratio of the strengths of the Rashba and Dresselhaus fields, and the direction of the beam changes by 90$^{\rm o}$ when the relative sign of the fields changes.   This surprising spatial anisotropy, originating from the anisotropic dispersion relations of electrons in the two fields, is due to general features of the energy contour surface of the electrons. Furthermore, the electron beam formation can be traced, using a stationary phase analysis of the real-space Green's function, to coalescing saddle points. Such beams should appear in two-contact transconductance\cite{Byers1995} as well as other transport\cite{Butler1985,Baranger1989} and scattering\cite{Crommie1993} phenomena.
A remnant of this anisotropy (although much weaker) appears to cause anisotropy in the dispersion relation of the spin\cite{Ullrich2003} and charge\cite{Badalyan2009} plasmon spectra for a quantum well with both Rashba and Dresselhaus fields.

The conduction band Hamiltonian describing an electron confined to a quantum well with [001] growth direction\cite{Dyakonov1986,Ivchenko1997}, to linear order in electron crystal momentum, is
\begin{equation} H = \frac{p^2}{2 m}  +\frac{\alpha}{\hbar} ( \sigma_x p_y - \sigma_y p_x) +\frac{\beta}{\hbar}(\sigma_x p_x -\sigma_y p_y ),  \label{H}
\end{equation}
where the first term is the kinetic energy, the second the Rashba interaction, and the third is the Dresselhaus interaction with strength $\beta$.    We reparameterize using
\begin{equation}
k_0   =  \frac{m}{\hbar^2}\sqrt{ \alpha^2 + \beta^2}, \qquad
\beta +i\alpha =  \frac{\hbar^2 k_0}{m} \exp(i\tau), 
 \end{equation}
yielding  $H = ({p^2}/{2 m}) + ({\hbar k_0}/{m}) U({\bf p})$,
where 
\begin{equation} 
U({\bf p}) = \sin(\tau) [\sigma_x p_y - \sigma_y p_x] +  \cos(\tau) [\sigma_x p_x -\sigma_y p_y].
\end{equation} 
 $U^2$ is proportional to the identity matrix:
\begin{equation}
U^2({\bf p}) = p^2 ( 1+ \sin(2\tau) \sin(2\theta_p) ) \equiv p^2 f_\tau(\theta_p), \label{U2}
\end{equation}
where $\theta_p$ is the angle that the momentum ${\bf p}$ makes with the positive $x$-axis. Although $U^2$ is diagonal in momentum and independent of spin, it is not isotropic.   

To calculate the electronic propagation we want  the retarded Green's function in coordinate space,
\begin{equation}
G_{\sigma,\sigma'}({\bf r}, {\bf r}') = \langle {\bf r},\sigma | \frac{1}{E - H + i\epsilon}|{\bf r}',\sigma' \rangle,
\end{equation}
where  $\epsilon \rightarrow 0^+.$
We obtain this from the Fourier transform of the momentum space Green's function $g_{\sigma,\sigma'}({\bf p},E)$,
\begin{equation}
G_{\sigma,\sigma'}({\bf r}, {\bf r}') = \frac{1}{(2 \pi \hbar)^2} \int d^2p e^{ i {\bf p} \cdot ({\bf r} -{\bf r}')/\hbar}
g_{\sigma, \sigma'}( {\bf p}; E).  \label{Green}
\end{equation}
Because $U^2$ is proportional to the $2 \times 2$ identity matrix,
the inverse of $E +i\epsilon -H$ can be found directly:
\begin{equation}
g(\hbar{\bf q}) = \frac{ (2m/\hbar^2)[ (k_E^2 - q^2) + 2 k_0 q U( \hat{{\bf q}})]}{q^4 -2 q^2(k_E^2 + 2 k_0^2 f_\tau) + k_E^4},\label{Denom}
\end{equation}
where  $k_E = ({ 2 m E/\hbar^2})^{1/2} $. 

\begin{figure}
\includegraphics[width=\columnwidth]{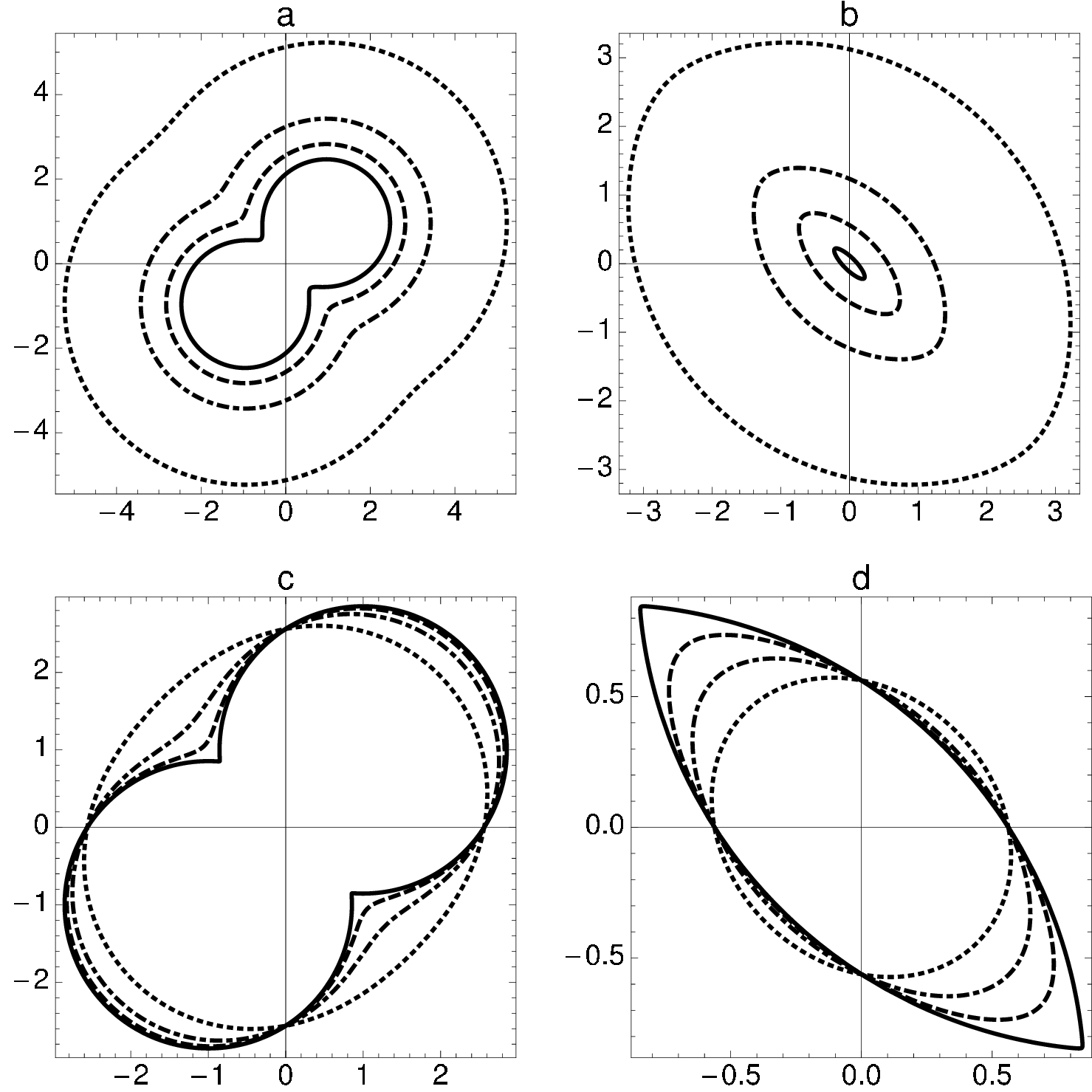}
\caption{Polar plots of $q_+(\theta_q,\tau)$  and 
$q_-(\theta_q,\tau)$, all momenta in units of $k_0$.   (a) $q_+,$  with $\tau = 0.971$ and $k_E= $ 0.5 (solid),  1.2 (dashed), 2. (dot-dashed), 4. (dotted).  (b) Same as (a) but plotting $q_-$. (c) $q_+$, $k_E = 1.2$, 
$\tau = $ 0.78 (solid), 0.971 (dashed), 1.15 (dot-dashed), 1.42 (dotted).  
(d) Same as (c) but plotting $q_-$ . Non-convexity for $q_+$ here causes the non-convexity in Fig. 4 and produces multiple stationary points. \label{energysurfaces} }
\end{figure}

The zeros of the denominator of the right-hand side of Eq.~\ref{Denom}, 
$q_\pm = | Q \pm  k_0 {f_\tau^{1/2}} | $,
with
$Q \equiv ({ k_E^2 + k_0^2 f_\tau})^{1/2},$
represent the intersection of the two energy surfaces $E_{\pm }({\bf q})$ with a constant energy plane, $E =({\hbar^2 k_E^2}/{2 m})$.  These intersections are shown in Fig.~\ref{energysurfaces}.  As the ratio of Rashba to Dresselhaus strengths changes (through $\tau$) or the electron energy changes (through $k_E$), the constant energy surfaces change from convex to non-convex, dramatically changing the angular dependence of the electron propagation.

Rewriting Eq.~(\ref{Denom}) using partial fractions,
\begin{flushleft}
$ G_{1,1}({\bf r}) = $ 
\begin{equation} 
   \frac{2 m}{(2 \pi \hbar)^2} \int_{\theta_r -\pi/2}^{\theta_r +\pi/2} d\theta_q 
\frac{-1}{Q}\left[ q_+ I_{1,1}( q_+ \rho) + q_- I_{1,1}(q_-\rho) \right]  \label{G11Angle}
\end{equation}
\end{flushleft}

$$G_{2,1}({\bf r}) =  
\frac{-2 mi}{(2 \pi \hbar)^2} \int_{\theta_r -\pi/2}^{\theta_r +\pi/2} d\theta_q 
\frac{ \sin(\theta_q +\tau) - i \cos( \theta_q -\tau)}{Q\sqrt{f_\tau}}$$
\begin{equation}\times\left[ q_+ I_{2,1}( q_+ \rho) - q_- I_{2,1}(q_-\rho) \right] , \label{G21Angle}
\end{equation}
where  $\rho = r \cos(\theta_q - \theta_r).$  In these equations the $I_{i,j}(z)$ are the radial integrals  ($\int p dp$) in the Fourier transform. Analytic expressions can be obtained for $I_{i,j}(z)$ (but are not shown as only the asymptotic form  $I_{l,m}(z) \propto\exp(iz)$ for large $z$ is relevant here).


\begin{figure}
\includegraphics[width=\columnwidth]{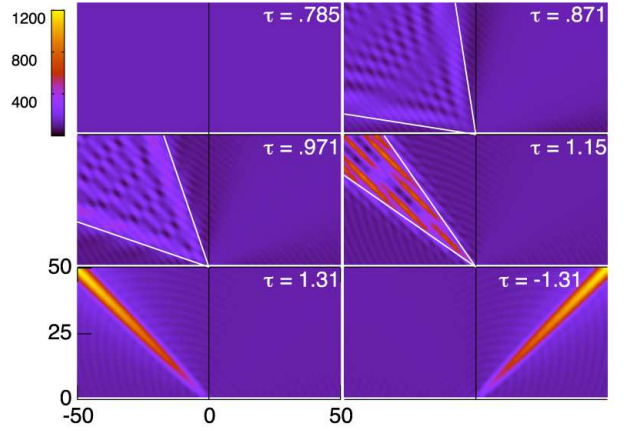}
\caption{ Scaled probability density for unpolarized spins $r  \sum_{i,j}^{1,2}|G_{i,j}({\bf r})|^2$ for various $\tau$ with positions in units of $k_0^{-1}$ and $k_E = 1.2k_0$.  In the upper left panel, $\tau \approx \pi/4$ (corresponding to $\alpha=\beta$) and the actual variation is 
less than 1\%.  In all panels the source point is (0,0) (bottom, center).
The white lines  mark the angle where two stationary points coalesce as discussed in the text. \label{probs}} 

\end{figure}

To highlight the angular anisotropy we plot the radial distance  times the probability density for {\it unpolarized} spins, $r  \sum_{i,j}^{1,2}|G_{i,j}({\bf r})|^2$, for $k_E = 1.2 k_o$, in Fig~\ref{probs}. Only the first two quadrants are shown since the function is $\pi$-periodic.  When $\tau = \pi/4$, corresponding to $|\alpha|=|\beta|$,  there is no anisotropy.   As $\tau$ is increased towards $\tau_{crit}(k_E)$, which is defined below ($\tau_{crit}(1.2k_0) = 1.31169$),
Fig.~\ref{probs} shows the electron probability pattern narrowing, rising and developing considerable structure, finally resolving into a narrow electron beam at $\tau=\tau_{crit}$. When 
$\tau = .871$, for example, there are interference patterns both in range and angle. The white lines mark a boundary between the variable, higher probability region and a more homogeneous region. The direction of these lines is computed below using stationary phase considerations.    When $\tau\approx\tau_{crit}$, the high probability region becomes very narrow and intense. Furthermore, along the direction $\theta_r=3\pi/4$, $r  \sum_{i,j}^{1,2}|G_{i,j}({\bf r})|^2$ increases  nearly monotonically. Finally when  $\tau =  - \tau_{crit}$, which corresponds to  changing the sign of the Rashba parameter $\alpha$ while leaving the Dresselhaus parameter $\beta$ unchanged, the beam is reoriented along the direction $\theta_r=\pi/4$. The angularly-integrated flux for all panels has been confirmed to be independent of $r$.

\begin{figure}
\includegraphics[width=\columnwidth]{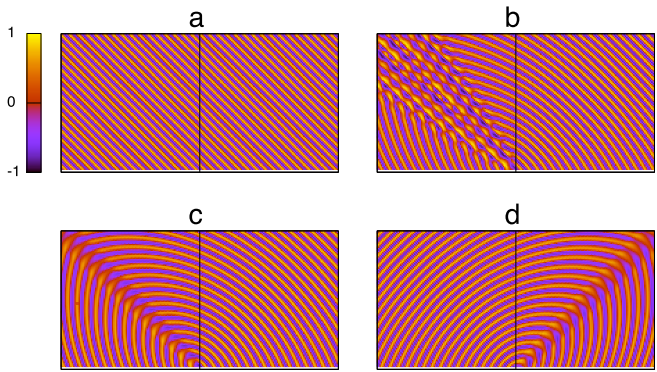}
\caption{ Polarization, Eq.~(\ref{polarization}), with positions in units of $k_0^{-1}$ and $k_E = 1.2k_0$. 
$\tau = $ (a) $\pi/4$, (b) $1.15$, (c) $1.31$  and (d) $-1.31$.\label{pols} }
 \end{figure}

Fig.~\ref{pols} shows the position-dependent polarization,
\begin{equation} 
P =    \frac{ |G_{1,1}({\bf r})|^2 -  |G_{2,1}({\bf r})|^2 }{ |G_{1,1}({\bf r})|^2 +  |G_{2,1}({\bf r})|^2},\label{polarization}
\end{equation}
of an injected spin polarized perpendicular to the quantum well plane, for  $\tau = \pi/4$, .871, $\tau_{crit}$, and $-\tau_{crit}$ when $k_E = 1.2k_0$.
The polarization changes dramatically as $\tau$ changes.  The polarization for $\tau =\pi/4$ suggests the spin helix described in Ref. \cite{Bernevig2006}.   The shift property $E_{- }({\bf q} + {\bf Q}) = E_{+ }({\bf q})$, used in Ref.  \cite{Bernevig2006}, yields an analytic result for the Green's functions in agreement with our results.

A stationary phase analysis of the angular  integrals in Eqs. \ref{G11Angle} and \ref{G21Angle}  provides (1) an explanation for why there is a region of high electron probability around the direction $\theta_r=3\pi/4$,  (2)  the angular width of these regions, and (3) the radial dependence of the electron probability along the direction $3\pi/4$ for $\tau=\tau_{crit}$.  It also provides a simple means of computing the electron probability at large $r$ when $\tau$ and $\theta_r$ are not too close to the white lines shown in Fig.~\ref{probs}.  

\begin{figure}
\includegraphics[width=\columnwidth]{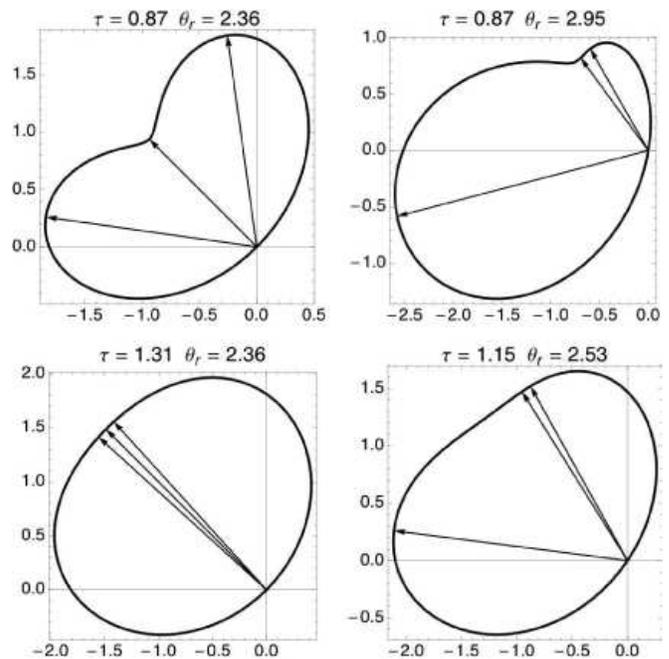}
\caption{Polar plots of the phase $\Phi_+(\theta_q, \tau,\theta_r) = q_+(\theta_q, \tau)\cos(\theta_q-\theta_r)$ with $\theta_r$ fixed and $\tau$ fixed. The right hand panels show two stationary points coalescing, for $\tau=0.871$ at $\theta_r $ about 35$^{\rm o}$ from $3 \pi /4$ and for $\tau=1.15$ at $\theta_r$ about 10$^{\rm o}$ from $3\pi/4$.   The arrows are drawn as normals to the curves and correspond the points of stationary phase. Similar plots of $\Phi_-$  (not shown) are  convex and have only one stationary point.\label{polarphase} } 
\end{figure}

The key result of stationary phase analysis is that the results seen in Fig.~\ref{probs} result from interfering contributions of stationary points, which can coalesce at critical values of the spin-orbit field or energy (through $\tau$ and $k_E$).
A typical integral in the computation is 
\begin{equation}
M_{1,1} = \int_{\theta_r - \pi/2}^{\theta_r + \pi/2} d\theta_q \frac{ q_+}{Q} I_{1,1}(q_+ \rho) \label{I11}
\end{equation} 
where  $\rho = r \cos(\theta_q - \theta_r).$ When  $r$ is large, the argument of $I_{1,1}$ will vary rapidly with the integration variable, $\theta_q$, and  dominant contributions to the integral will originate at points of stationary phase, where the derivative of the argument of $I_{1,1}$ with respect to $\theta_q$ vanishes.  
Apart from the endpoints, the argument of $I_{1,1}$ is positive throughout the chosen range of integration. Following 
Ref.~\cite{Bleistein1975}, $I_{1,1}$ can be replaced by its large-argument asymptotic form $\propto\exp(iz)$.
With this replacement, $M_{1,1}$ becomes
$$
M_{1,1} \sim \int  d\theta_q  \frac{i\pi q_+}{2Q}\exp[iq_+ r\cos(\theta_q - \theta_r)].
$$
Figure~\ref{polarphase} shows polar plots of $\Phi_+ =q_+\cos(\theta -\theta_r).$ Depending on $\tau$ and $\theta_r$ the phase
$\Phi_+$ can exhibit either one or three points of stationary phase. The arrows in Fig.~\ref{polarphase} are 
drawn along radials which are perpendicular to the curves, which 
correspond to points of stationary phase.  Comparing the polar plots of Fig.~\ref{energysurfaces} and Fig.~\ref{polarphase} it is apparent that multiple stationary points arise from the non-convexity of the  $q_+$ curve. In contrast the 
$q_-$ curve is always convex and therefore  the phase $\Phi(q_-)$ always shows only a single
stationary point.   The right panels in Fig.~\ref{polarphase} correspond to observation angles ($\theta_r$) such that two stationary points are about to merge and disappear leaving only one stationary point. These observation angles correspond to the white lines in Fig.~\ref{probs} and give a rough estimate of the size of the intense areas there. {\it All three stationary phase points merge at a value of $\tau=\tau_{crit}$}.  The lower left panel in Fig.~\ref{polarphase} shows $\Phi_+$ for $\tau$ near $\tau_{crit}$, corresponding to the narrowest beams in Fig.~\ref{probs}.

At $\tau=\tau_{crit}$  the phase $\Phi_+\propto (\theta_q - 3\pi/4)^4$.  A simple stationary phase analysis fails,  but the quartic behavior implies that $M_{1,1}\propto r^{-1/4} ,$ which is consistent with the electron probability 
times $r$ increasing as $r^{1/2}$. A radial plot (not shown here) of the probability of Fig.~\ref{probs} along $3\pi/4$ shows this behavior.   Thus the stationary phase analysis accounts for the angular widths of the intense regions of  Fig.~\ref{probs}; they are determined by the transition from three points to one point of stationary phase. In turn this is where both the first and second derivatives of the phase vanish, which is the criterion for the white lines  in Fig.~\ref{probs}. We have verified that the angular integral of the radial component of the flux times $r$ is in fact independent of $r$.Ê A plot (in EPAPS) of the radial component of the flux times $r$ as a function of $\theta_r$ shows that the peak flux times $r$ rises proportionally to $r^{1/2}$ while at the same time the peak narrows, as it must for the total integral to be fixed.
The case of $\tau = \pi/4$, corresponding to $\alpha=\beta$ is a singular case for which the stationary phase arguments here do not apply. 

In the smooth regions of Fig~\ref{probs}, it can be shown that there is only a single point of stationary phase associated with $\Phi_+$ .   This must be combined with the single point of stationary phase associated with $\Phi_-$, and the interference of these terms accounts for spin precession and the periodicity visible in the polarization plots of Fig.~\ref{probs} just beyond the regions of enhanced  probability.  


We conclude with a discussion of length scales for the phenomenon of ``electron beams'' and the implications for other measurements in quantum wells. In all our results, $r$, has been shown  in units of $1/k_0$.  Experimental values of $\alpha$ range from  $\sim 0.1-0.3$~eV\AA\ in InGaAs quantum wells\cite{Nitta1997,Sato2001} and $\sim0.5$~eV\AA\ in InAs quantum wells\cite{Grundler2000}, with $\beta$ of the same order. Using the appropriate effective masses ($\sim 0.05m_o$ for InGaAs, $\sim 0.025m_o$ for InAs, where $m_o$ is the free mass) this produces a length scale of 
$1/k_0 \approx 40 \mbox{nm}$.  Thus the scale of the plots, Figs.~\ref{probs} and~\ref{pols}, is about $2\mu$m. It may be possible to image these patterns, as in Ref.~\cite{Topinka2001}. However, the dramatic enhancements, shown over a longer length scale for clarity, also occur on smaller scales $\sim 100$~nm.  For GaAs/AlGaAs $\alpha$ is an order of magnitude smaller, meaning the scale of Figs.~\ref{probs} and~\ref{pols} would be $20\mu$m, however in this system the mean free paths frequently exceed $100\mu$m\cite{Umansky1996}. 

Fig.~\ref{probs}'s results could be measured directly by two-contact transconductance\cite{Byers1995}, however the highly anisotropic electron propagation dramatically affects many other properties.  The local density of states near impurities has spatial structure described by the square of the position-dependent Green's function\cite{Crommie1993}, and thus the wave functions of impurities seen in scanning tunneling microscopy should be highly anisotropic. Scattering of electrons from impurities will also be highly anisotropic, yielding correspondingly anisotropic diffusive transport [in a real-space Kubo formalism for diffusive transport\cite{Butler1985,Baranger1989}, the real-space Green's functions appear in similar combination to Fig.~\ref{probs}, with a spatial derivative selecting the conductivity direction ($\sigma_{\alpha\beta} \propto \int d{\bf r} [\nabla_{\alpha}G({\bf r})][\nabla_{\beta}G({\bf r})]$)].  These derivatives of the Green's function have the same spatial structure as shown in Fig.~\ref{probs}.

We have shown that highly anisotropic electron propagation (electron beams) occurs in a semiconductor quantum well at appropriate relative values of the Dresselhaus and Rashba spin-orbit fields. In addition to the fundamental consequences for properties of the quantum well, driven by the peculiar electronic structure, this system offers the intriguing possibility of altering and even redirecting the narrow electron beams of Fig.~\ref{probs} by 90$^{\rm o}$ by reversing the sign of the Rashba field ($\alpha$), by varying an electric field perpendicular  to the quantum well. 

We acknowledge support from NRI, an ONR MURI and helpful discussions with B. Moehlmann.


\begin{thebibliography}{38}
\expandafter\ifx\csname natexlab\endcsname\relax\def\natexlab#1{#1}\fi
\expandafter\ifx\csname bibnamefont\endcsname\relax
  \def\bibnamefont#1{#1}\fi
\expandafter\ifx\csname bibfnamefont\endcsname\relax
  \def\bibfnamefont#1{#1}\fi
\expandafter\ifx\csname citenamefont\endcsname\relax
  \def\citenamefont#1{#1}\fi
\expandafter\ifx\csname url\endcsname\relax
  \def\url#1{\texttt{#1}}\fi
\expandafter\ifx\csname urlprefix\endcsname\relax\def\urlprefix{URL }\fi
\providecommand{\bibinfo}[2]{#2}
\providecommand{\eprint}[2][]{\url{#2}}

\bibitem[{\citenamefont{Wolf et~al.}(2001)}]{Wolf2001}
\bibinfo{author}{\bibfnamefont{S.~A.} \bibnamefont{Wolf}} \bibnamefont{et~al.},
  \bibinfo{journal}{Science} \textbf{\bibinfo{volume}{294}},
  \bibinfo{pages}{1488} (\bibinfo{year}{2001}).

\bibitem[{\citenamefont{Awschalom et~al.}(2002)\citenamefont{Awschalom,
  Samarth, and Loss}}]{Awschalom2002}
\bibinfo{editor}{\bibfnamefont{D.~D.} \bibnamefont{Awschalom}},
  \bibinfo{editor}{\bibfnamefont{N.}~\bibnamefont{Samarth}}, \bibnamefont{and}
  \bibinfo{editor}{\bibfnamefont{D.}~\bibnamefont{Loss}}, eds.,
  \emph{\bibinfo{title}{Semiconductor Spintronics and Quantum Computation}}
  (\bibinfo{publisher}{Springer Verlag}, \bibinfo{address}{Heidelberg},
  \bibinfo{year}{2002}).

\bibitem[{\citenamefont{Ziese and Thornton}(2001)}]{Ziese2001}
\bibinfo{editor}{\bibfnamefont{M.}~\bibnamefont{Ziese}} \bibnamefont{and}
  \bibinfo{editor}{\bibfnamefont{M.~J.} \bibnamefont{Thornton}}, eds.,
  \emph{\bibinfo{title}{Spin electronics}}, vol. \bibinfo{volume}{569} of
  \emph{\bibinfo{series}{Lecture Notes in Physics}}
  (\bibinfo{publisher}{Springer-Verlag}, \bibinfo{address}{Heidelberg},
  \bibinfo{year}{2001}).

\bibitem[{\citenamefont{Awschalom and Flatt\'e}(2007)}]{Awschalom2007}
\bibinfo{author}{\bibfnamefont{D.~D.} \bibnamefont{Awschalom}}
  \bibnamefont{and} \bibinfo{author}{\bibfnamefont{M.~E.}
  \bibnamefont{Flatt\'e}}, \bibinfo{journal}{Nature Physics}
  \textbf{\bibinfo{volume}{3}}, \bibinfo{pages}{153} (\bibinfo{year}{2007}).

\bibitem[{\citenamefont{Schliemann et~al.}(2003)\citenamefont{Schliemann,
  Egues, and Loss}}]{Schliemann2003}
\bibinfo{author}{\bibfnamefont{J.}~\bibnamefont{Schliemann}},
  \bibinfo{author}{\bibfnamefont{J.~C.} \bibnamefont{Egues}}, \bibnamefont{and}
  \bibinfo{author}{\bibfnamefont{D.}~\bibnamefont{Loss}},
  \bibinfo{journal}{\prl} \textbf{\bibinfo{volume}{90}},
  \bibinfo{pages}{146801} (\bibinfo{year}{2003}).

\bibitem[{\citenamefont{Kato et~al.}(2004{\natexlab{a}})\citenamefont{Kato,
  Myers, Gossard, and Awschalom}}]{Kato2004c}
\bibinfo{author}{\bibfnamefont{Y.~K.} \bibnamefont{Kato}},
  \bibinfo{author}{\bibfnamefont{R.~C.} \bibnamefont{Myers}},
  \bibinfo{author}{\bibfnamefont{A.~C.} \bibnamefont{Gossard}},
  \bibnamefont{and} \bibinfo{author}{\bibfnamefont{D.~D.}
  \bibnamefont{Awschalom}}, \bibinfo{journal}{Nature}
  \textbf{\bibinfo{volume}{427}}, \bibinfo{pages}{50}
  (\bibinfo{year}{2004}{\natexlab{a}}).

\bibitem[{\citenamefont{Crooker and Smith}(2005)}]{Crooker2005a}
\bibinfo{author}{\bibfnamefont{S.~A.} \bibnamefont{Crooker}} \bibnamefont{and}
  \bibinfo{author}{\bibfnamefont{D.~L.} \bibnamefont{Smith}},
  \bibinfo{journal}{\prl} \textbf{\bibinfo{volume}{94}},
  \bibinfo{pages}{236601} (\bibinfo{year}{2005}).

\bibitem[{\citenamefont{Crooker et~al.}(2005)}]{Crooker2005b}
\bibinfo{author}{\bibfnamefont{S.~A.} \bibnamefont{Crooker}}
  \bibnamefont{et~al.}, \bibinfo{journal}{Science}
  \textbf{\bibinfo{volume}{309}}, \bibinfo{pages}{2191} (\bibinfo{year}{2005}).

\bibitem[{\citenamefont{Bernevig et~al.}(2006)\citenamefont{Bernevig,
  Orenstein, and Zhang}}]{Bernevig2006}
\bibinfo{author}{\bibfnamefont{B.~A.} \bibnamefont{Bernevig}},
  \bibinfo{author}{\bibfnamefont{J.}~\bibnamefont{Orenstein}},
  \bibnamefont{and} \bibinfo{author}{\bibfnamefont{S.-C.} \bibnamefont{Zhang}},
  \bibinfo{journal}{Physical Review Letters} \textbf{\bibinfo{volume}{97}},
  \bibinfo{eid}{236601} (\bibinfo{year}{2006}).

\bibitem[{\citenamefont{Koralek et~al.}(2009)}]{Koralek2009}
\bibinfo{author}{\bibfnamefont{J.~D.} \bibnamefont{Koralek}}
  \bibnamefont{et~al.}, \bibinfo{journal}{Nature}
  \textbf{\bibinfo{volume}{458}}, \bibinfo{pages}{610} (\bibinfo{year}{2009}).

\bibitem[{\citenamefont{Edelstein}(1990)}]{Edelstein1990}
\bibinfo{author}{\bibfnamefont{V.~M.} \bibnamefont{Edelstein}},
  \bibinfo{journal}{Solid State Comm.} \textbf{\bibinfo{volume}{73}},
  \bibinfo{pages}{233} (\bibinfo{year}{1990}).

\bibitem[{\citenamefont{D'yakonov and Perel'}(1971)}]{Dyakonov1971}
\bibinfo{author}{\bibfnamefont{M.~I.} \bibnamefont{D'yakonov}}
  \bibnamefont{and} \bibinfo{author}{\bibfnamefont{V.~I.}
  \bibnamefont{Perel'}}, \bibinfo{journal}{Physics Letters A}
  \textbf{\bibinfo{volume}{35}}, \bibinfo{pages}{459} (\bibinfo{year}{1971}).

\bibitem[{\citenamefont{Hirsch}(1999)}]{Hirsch1999}
\bibinfo{author}{\bibfnamefont{J.~E.} \bibnamefont{Hirsch}},
  \bibinfo{journal}{\prl} \textbf{\bibinfo{volume}{83}}, \bibinfo{pages}{1834}
  (\bibinfo{year}{1999}).

\bibitem[{\citenamefont{Murakami et~al.}(2003)\citenamefont{Murakami, Nagaosa,
  and Zhang}}]{Murakami2003}
\bibinfo{author}{\bibfnamefont{S.}~\bibnamefont{Murakami}},
  \bibinfo{author}{\bibfnamefont{N.}~\bibnamefont{Nagaosa}}, \bibnamefont{and}
  \bibinfo{author}{\bibfnamefont{S.-C.} \bibnamefont{Zhang}},
  \bibinfo{journal}{Science} \textbf{\bibinfo{volume}{301}},
  \bibinfo{pages}{1348} (\bibinfo{year}{2003}).

\bibitem[{\citenamefont{Sinova et~al.}(2004)}]{Sinova2004}
\bibinfo{author}{\bibfnamefont{J.}~\bibnamefont{Sinova}} \bibnamefont{et~al.},
  \bibinfo{journal}{\prl} \textbf{\bibinfo{volume}{92}},
  \bibinfo{pages}{126603} (\bibinfo{year}{2004}).

\bibitem[{\citenamefont{Kato et~al.}(2004{\natexlab{b}})\citenamefont{Kato,
  Myers, Gossard, and Awschalom}}]{Kato2004b}
\bibinfo{author}{\bibfnamefont{Y.~K.} \bibnamefont{Kato}},
  \bibinfo{author}{\bibfnamefont{R.~C.} \bibnamefont{Myers}},
  \bibinfo{author}{\bibfnamefont{A.~C.} \bibnamefont{Gossard}},
  \bibnamefont{and} \bibinfo{author}{\bibfnamefont{D.~D.}
  \bibnamefont{Awschalom}}, \bibinfo{journal}{Science}
  \textbf{\bibinfo{volume}{306}}, \bibinfo{pages}{1910}
  (\bibinfo{year}{2004}{\natexlab{b}}).

\bibitem[{\citenamefont{Walls et~al.}(2006)\citenamefont{Walls, Huang,
  Westervelt, and Heller}}]{Walls2006}
\bibinfo{author}{\bibfnamefont{J.~D.} \bibnamefont{Walls}},
  \bibinfo{author}{\bibfnamefont{J.}~\bibnamefont{Huang}},
  \bibinfo{author}{\bibfnamefont{R.~M.} \bibnamefont{Westervelt}},
  \bibnamefont{and} \bibinfo{author}{\bibfnamefont{E.~J.}
  \bibnamefont{Heller}}, \bibinfo{journal}{Phys. Rev. B}
  \textbf{\bibinfo{volume}{73}}, \bibinfo{eid}{035325} (\bibinfo{year}{2006}).

\bibitem[{\citenamefont{Csord\'as et~al.}(2006)\citenamefont{Csord\'as, Cserti,
  P\'alyi, and Z\"ulicke}}]{Csordas2006}
\bibinfo{author}{\bibfnamefont{A.}~\bibnamefont{Csord\'as}},
  \bibinfo{author}{\bibfnamefont{J.}~\bibnamefont{Cserti}},
  \bibinfo{author}{\bibfnamefont{A.}~\bibnamefont{P\'alyi}}, \bibnamefont{and}
  \bibinfo{author}{\bibfnamefont{U.}~\bibnamefont{Z\"ulicke}},
  \bibinfo{journal}{Eur. Phys. J. B} \textbf{\bibinfo{volume}{54}},
  \bibinfo{pages}{189} (\bibinfo{year}{2006}).

\bibitem[{\citenamefont{Rashba}(1960)}]{Rashba1960}
\bibinfo{author}{\bibfnamefont{E.~I.} \bibnamefont{Rashba}},
  \bibinfo{journal}{Soviet Physics Solid State} \textbf{\bibinfo{volume}{2}},
  \bibinfo{pages}{1109} (\bibinfo{year}{1960}).

\bibitem[{\citenamefont{Bychkov and Rashba}(1984)}]{Bychkov1984}
\bibinfo{author}{\bibfnamefont{Y.~A.} \bibnamefont{Bychkov}} \bibnamefont{and}
  \bibinfo{author}{\bibfnamefont{E.~I.} \bibnamefont{Rashba}},
  \bibinfo{journal}{J. Phys. C} \textbf{\bibinfo{volume}{17}},
  \bibinfo{pages}{6039} (\bibinfo{year}{1984}).

\bibitem[{\citenamefont{Br\"uning et~al.}(2007)\citenamefont{Br\"uning, Geyler,
  and Pankrashkin}}]{Bruning2007}
\bibinfo{author}{\bibfnamefont{J.}~\bibnamefont{Br\"uning}},
  \bibinfo{author}{\bibfnamefont{V.}~\bibnamefont{Geyler}}, \bibnamefont{and}
  \bibinfo{author}{\bibfnamefont{K.}~\bibnamefont{Pankrashkin}},
  \bibinfo{journal}{J. Phys. A} \textbf{\bibinfo{volume}{40}},
  \bibinfo{pages}{F697} (\bibinfo{year}{2007}).

\bibitem[{\citenamefont{Dresselhaus}(1955)}]{Dresselhaus1955}
\bibinfo{author}{\bibfnamefont{G.}~\bibnamefont{Dresselhaus}},
  \bibinfo{journal}{\pr} \textbf{\bibinfo{volume}{100}}, \bibinfo{pages}{580}
  (\bibinfo{year}{1955}).

\bibitem[{\citenamefont{Trushin and Schliemann}(2007)}]{Trushin2007}
\bibinfo{author}{\bibfnamefont{M.}~\bibnamefont{Trushin}} \bibnamefont{and}
  \bibinfo{author}{\bibfnamefont{J.}~\bibnamefont{Schliemann}},
  \bibinfo{journal}{Phys. Rev. B} \textbf{\bibinfo{volume}{75}},
  \bibinfo{pages}{155323} (\bibinfo{year}{2007}).

\bibitem[{\citenamefont{Trushin et~al.}(2009)}]{Trushin2009}
\bibinfo{author}{\bibfnamefont{M.}~\bibnamefont{Trushin}} \bibnamefont{et~al.},
  \bibinfo{journal}{Phys. Rev. B} \textbf{\bibinfo{volume}{80}},
  \bibinfo{pages}{134405} (\bibinfo{year}{2009}).

\bibitem[{\citenamefont{Byers and Flatt\'e}(1995)}]{Byers1995}
\bibinfo{author}{\bibfnamefont{J.~M.} \bibnamefont{Byers}} \bibnamefont{and}
  \bibinfo{author}{\bibfnamefont{M.~E.} \bibnamefont{Flatt\'e}},
  \bibinfo{journal}{Phys. Rev. Lett.} \textbf{\bibinfo{volume}{74}},
  \bibinfo{pages}{306} (\bibinfo{year}{1995}).

\bibitem[{\citenamefont{Butler}(1985)}]{Butler1985}
\bibinfo{author}{\bibfnamefont{W.~H.} \bibnamefont{Butler}},
  \bibinfo{journal}{Phys. Rev. B} \textbf{\bibinfo{volume}{31}},
  \bibinfo{pages}{3260} (\bibinfo{year}{1985}).

\bibitem[{\citenamefont{Baranger and Stone}(1989)}]{Baranger1989}
\bibinfo{author}{\bibfnamefont{H.~U.} \bibnamefont{Baranger}} \bibnamefont{and}
  \bibinfo{author}{\bibfnamefont{A.~D.} \bibnamefont{Stone}},
  \bibinfo{journal}{\prb} \textbf{\bibinfo{volume}{40}}, \bibinfo{pages}{8169}
  (\bibinfo{year}{1989}).

\bibitem[{\citenamefont{Crommie et~al.}(1993)\citenamefont{Crommie, Lutz, and
  Eigler}}]{Crommie1993}
\bibinfo{author}{\bibfnamefont{M.~F.} \bibnamefont{Crommie}},
  \bibinfo{author}{\bibfnamefont{C.~P.} \bibnamefont{Lutz}}, \bibnamefont{and}
  \bibinfo{author}{\bibfnamefont{D.~M.} \bibnamefont{Eigler}},
  \bibinfo{journal}{Nature} \textbf{\bibinfo{volume}{363}},
  \bibinfo{pages}{524} (\bibinfo{year}{1993}).

\bibitem[{\citenamefont{Ullrich and Flatt\'e}(2003)}]{Ullrich2003}
\bibinfo{author}{\bibfnamefont{C.~A.} \bibnamefont{Ullrich}} \bibnamefont{and}
  \bibinfo{author}{\bibfnamefont{M.~E.} \bibnamefont{Flatt\'e}},
  \bibinfo{journal}{Phys. Rev. B} \textbf{\bibinfo{volume}{68}},
  \bibinfo{pages}{235310} (\bibinfo{year}{2003}).

\bibitem[{\citenamefont{Badalyan et~al.}(2009)\citenamefont{Badalyan,
  Matos-Abiague, Vignale, and Fabian}}]{Badalyan2009}
\bibinfo{author}{\bibfnamefont{S.~M.} \bibnamefont{Badalyan}},
  \bibinfo{author}{\bibfnamefont{A.}~\bibnamefont{Matos-Abiague}},
  \bibinfo{author}{\bibfnamefont{G.}~\bibnamefont{Vignale}}, \bibnamefont{and}
  \bibinfo{author}{\bibfnamefont{J.}~\bibnamefont{Fabian}},
  \bibinfo{journal}{Phys. Rev. B} \textbf{\bibinfo{volume}{79}},
  \bibinfo{eid}{205305} (\bibinfo{year}{2009}).

\bibitem[{\citenamefont{D'yakonov and Kachorovskii}(1986)}]{Dyakonov1986}
\bibinfo{author}{\bibfnamefont{M.~I.} \bibnamefont{D'yakonov}}
  \bibnamefont{and} \bibinfo{author}{\bibfnamefont{V.~Y.}
  \bibnamefont{Kachorovskii}}, \bibinfo{journal}{Soviet Physics Semiconductors}
  \textbf{\bibinfo{volume}{20}}, \bibinfo{pages}{110} (\bibinfo{year}{1986}).

\bibitem[{\citenamefont{Ivchenko and Pikus}(1997)}]{Ivchenko1997}
\bibinfo{author}{\bibfnamefont{E.~L.} \bibnamefont{Ivchenko}} \bibnamefont{and}
  \bibinfo{author}{\bibfnamefont{G.~E.} \bibnamefont{Pikus}},
  \emph{\bibinfo{title}{Superlattices and Other Heterostructures}}
  (\bibinfo{publisher}{Springer}, \bibinfo{address}{New York},
  \bibinfo{year}{1997}).

\bibitem[{\citenamefont{Bleistein and Handelsman}(1975)}]{Bleistein1975}
\bibinfo{author}{\bibfnamefont{N.}~\bibnamefont{Bleistein}} \bibnamefont{and}
  \bibinfo{author}{\bibfnamefont{R.~A.} \bibnamefont{Handelsman}},
  \emph{\bibinfo{title}{Asymptotic expansions of integrals}}
  (\bibinfo{publisher}{Holt, Rinhart and Winston}, \bibinfo{address}{New York},
  \bibinfo{year}{1975}).

\bibitem[{\citenamefont{Nitta et~al.}(1997)\citenamefont{Nitta, Akazaki,
  Takayanagi, and Enoki}}]{Nitta1997}
\bibinfo{author}{\bibfnamefont{J.}~\bibnamefont{Nitta}},
  \bibinfo{author}{\bibfnamefont{T.}~\bibnamefont{Akazaki}},
  \bibinfo{author}{\bibfnamefont{H.}~\bibnamefont{Takayanagi}},
  \bibnamefont{and} \bibinfo{author}{\bibfnamefont{T.}~\bibnamefont{Enoki}},
  \bibinfo{journal}{Phys. Rev. Lett.} \textbf{\bibinfo{volume}{78}},
  \bibinfo{pages}{1335} (\bibinfo{year}{1997}).

\bibitem[{\citenamefont{Sato et~al.}(2001)\citenamefont{Sato, Kita, Gozu, and
  Yamada}}]{Sato2001}
\bibinfo{author}{\bibfnamefont{Y.}~\bibnamefont{Sato}},
  \bibinfo{author}{\bibfnamefont{T.}~\bibnamefont{Kita}},
  \bibinfo{author}{\bibfnamefont{S.}~\bibnamefont{Gozu}}, \bibnamefont{and}
  \bibinfo{author}{\bibfnamefont{S.}~\bibnamefont{Yamada}},
  \bibinfo{journal}{Journal of Applied Physics} \textbf{\bibinfo{volume}{89}},
  \bibinfo{pages}{8017} (\bibinfo{year}{2001}).

\bibitem[{\citenamefont{Grundler}(2000)}]{Grundler2000}
\bibinfo{author}{\bibfnamefont{D.}~\bibnamefont{Grundler}},
  \bibinfo{journal}{Phys. Rev. Lett.} \textbf{\bibinfo{volume}{84}},
  \bibinfo{pages}{6074} (\bibinfo{year}{2000}).

\bibitem[{\citenamefont{Topinka et~al.}(2001)}]{Topinka2001}
\bibinfo{author}{\bibfnamefont{M.~A.} \bibnamefont{Topinka}}
  \bibnamefont{et~al.}, \bibinfo{journal}{Nature}
  \textbf{\bibinfo{volume}{410}}, \bibinfo{pages}{183} (\bibinfo{year}{2001}).

\bibitem[{\citenamefont{Umansky et~al.}(1996)\citenamefont{Umansky,
  de~Picciotto, and Heiblum}}]{Umansky1996}
\bibinfo{author}{\bibfnamefont{V.}~\bibnamefont{Umansky}},
  \bibinfo{author}{\bibfnamefont{R.}~\bibnamefont{de~Picciotto}},
  \bibnamefont{and} \bibinfo{author}{\bibfnamefont{M.}~\bibnamefont{Heiblum}},
  \bibinfo{journal}{\apl} \textbf{\bibinfo{volume}{71}}, \bibinfo{pages}{683}
  (\bibinfo{year}{1996}).

\end{thebibliography}

\end{document}